# Cancer gene prioritization by integrative analysis of mRNA expression and DNA copy number data: a comparative review

Leo Lahti (1,2), Martin Schäfer (3), Hans-Ulrich Klein (4), Silvio Bicciato (5) and Martin Dugas (4)

(1) Aalto University, Department of Information and Computer Science, Adaptive Informatics Research Centre and Helsinki Institute for Information Technology HIIT, P.O. Box 15400, FI-00076 Aalto, Finland, (2) University of Helsinki, Department of Veterinary Bioscience, P.O. Box 66, FI-00014, Finland. (3) Department of Statistics, TU Dortmund University, Vogelpothsweg 87, 44221 Dortmund, Germany (4) Institute of Medical Informatics, University of Münster, Albert-Schweitzer-Campus 1 - A11, D-48149 Münster, Germany (5) Center for Genome Research, University of Modena and Reggio Emilia, via G. Campi, 287 - 41100 Modena, Italy

***Abstract***

*A variety of genome-wide profiling techniques are available to probe complementary aspects of genome structure and function. Integrative analysis of heterogeneous data sources can reveal higher-level interactions that cannot be detected based on individual observations. A standard integration task in cancer studies is to identify altered genomic regions that induce changes in the expression of the associated genes based on joint analysis of genome-wide gene expression and copy number profiling measurements. In this review, we provide a comparison among various modeling procedures for integrating genome-wide profiling data of gene copy number and transcriptional alterations and highlight common approaches to genomic data integration. A transparent benchmarking procedure is introduced to quantitatively compare the cancer gene prioritization performance of the alternative methods. The benchmarking algorithms and data sets are available at* http://intcomp.r-forge.r-project.org

***Keywords:*** *algorithms, cancer, data integration, DNA copy number, gene expression, microarrays*

## INTRODUCTION

Genome-wide profiling technologies, in particular microarrays and next generation sequencing, are used to characterize disease-associated changes at various levels of genome function. Identification of the key players - genes, chromosomal regions, or biological processes - is a fundamental step towards mechanistic characterization of the disease and revealing molecular targets for potential therapeutic intervention. Genomic, transcriptomic, epigenomic, and proteomic measurements characterize different aspects of genome regulation and function that are particularly relevant for cancer research [1-2]. Integrative analysis has been used to prioritize disease genes or chromosomal regions for experimental testing, to discover disease subtypes [3-4], or to predict patient survival or other clinical variables [5]. Co-occurring genomic observations are increasingly available in private and public repositories, such as the Cancer Genome Atlas database [6] and the Leukemia Gene Atlas [7], which promotes wide access to data resources. However, the lack of algorithmic implementations still represents a bottleneck hampering integrative approaches.

The integration of gene expression (GE) and copy number (CN) data to identify DNA copy number alterations that induce changes in the expression levels of the associated genes is a common task in cancer studies [8], and the detection of chromosomal regions with exceptionally high statistical association between CN and GE can pinpoint disease genes and potential cancer mechanisms [9-10]. First high-throughput analyses were reported about a decade ago [11-13], evidencing a clear cis-dosage effect of CN alterations on GE levels [14-16]. Although the downstream effect of CN alteration on gene expression is still a focus of ongoing research [17-18], a systematic quantitative comparison of alternative approaches for integrating GE/CN data sets has been missing, as clearly highlighted by the recent review by Huang *et al* [8]. Hence, we designed a transparent and quantitative benchmarking procedure to compare publicly available methods for cancer gene prioritization from the integrative analysis of CN/GE profiling data. This benchmark approach was applied to assess and compare the performance of 12 algorithms on two simulated data sets and three real case studies. In the following sections, we overview available methodologies for cancer gene prioritization based on GE/CN data integration, introduce the analysis pipeline, and discuss benchmarking results.

## Quantifying associations between gene expression and copy number

**Two-step approaches** Several approaches separately assess the alterations in each data set and compare results from CN and GE analyses to detect simultaneous changes, mostly modeling changes in GE based on the CN signals [16,19]. This corresponds to the biological intuition concerning the cis-regulatory effect of CN alterations. In the first step samples are grouped based on hard CN calls, call probabilities [20] or quantiles (DR-SAM [21]). In the second step, differential gene expression is quantified between such groups using e.g., standard approaches for GE data analysis as t-test [13], nonparametric alternatives [22] or permutation tests [23-24]. Some methods focus on regions rather than probes/genes [19, 24]. Comparison of gene expression levels between the sample groups with distinct copy number status on a particular chromosomal region is aimed to reveal copy number-induced transcriptional responses, typically within the affected region.

**Regression approaches** Another class of tools uses regression models with CN as the predictor and GE as the response variable, again following biological intuition concerning the cis-regulatory effect of CN alterations. Both linear [12] and non-linear regression models [25] have been proposed. Regression models have been designed both based on one-to-one correspondence between CN and GE probes [26] and multiple/multivariate linear regression [14, 26-28].

**Correlation-based approaches** The DR-Correlate [21] and a modified version of Ortiz-Estevez algorithm [16] use correlation-based analysis to scan over the genome to detect loci with exceptionally high associations between CN/GE. Schäfer et al. [29] substitute the sample means by the reference medians, and Lipson et al. [30] use quantile-based analysis to obtain improved correlation coefficients. Canonical correlation analysis (CCA) identifies linear combinations of CN and GE probes that are maximally correlated. Various modifications for dimensionality reduction and model regularization have been proposed based on principal component analysis [31] and penalized approaches based, e.g., on LASSO or elastic net to obtain sparse versions of CCA [5, 32-33], or based on variants that focus on specific types of dependency [34]. Regularization can reduce overfitting, and sparsity can simplify interpretation of the results, but determining appropriate regularization parameters may represent a challenging issue.

**Latent variable models** Latent variable approaches model directly the data-generating processes and noise. The pint/simcca [34] decomposes GE and CN data sets into shared and independent Gaussian components based on regularized probabilistic CCA. The algorithm by [4] is a related model suited for sample classification and subtype discovery. Latent matrix decomposition models and iterative, dependency-seeking projections have also been suggested based on generalized singular-value decomposition [3] and independent component analysis [35].

## Benchmarking the algorithms

Manual literature search in PubMed and Google Scholar using combinations of the keywords ‘gene expression’, ‘copy number’, ‘integration’, and curation of the Bioconductor repository (http://www.bioconductor.org) were performed to identify available implementations, yielding 12 algorithms that were applicable for cancer gene prioritization based on integrative analysis of GE/CN data (Table 1). The source code for Ortiz-Estevez [16] was obtained from the authors. An automated benchmarking pipeline (‘intcomp’) was created to compare method performance on two simulated data sets and three real case studies. Calculations were carried out in R (2.13.2 [36] and intcomp v. 0.3.27). The comparison pipeline is available through R-Forge (http://intcomp.r-forge.r-project.org/) and the algorithm versions are detailed in the package vignette.

Each method was used to prioritize candidate cancer genes, followed by comparison to a golden standard list of known cancer genes, and ranking of the methods based on (i) true positive rate among the top findings, (ii) receiver operating characteristic (ROC) analysis of the overall prioritized gene list, and (iii) running times. Since only a subset of genes are likely to be cancer-associated, the standard Area Under Curve (AUC) analysis, which considers the overall prioritized gene list, was complemented by investigating the true positive rate among the top findings. This is more appropriate in particular for methods such as CNAmet, Ortiz-Estevez, or PREDA/SODEGIR, which have originally been designed to detect altered chromosomal regions rather than to prioritize individual genes. Default parameters for each method were used where possible. The following exceptions were made to apply the algorithms to cancer gene prioritization. In DR-Correlate [21], empirical p-values from 1000 random gene permutations were used to rank the genes. The DR-Correlate t-test option was not applicable on the Ferrari simulations due to the low number of replicate samples. CNAmet [24, 37] requires called CN values and provides separate lists for amplifications and deletions; thus, the two lists were pooled and ranked based on the p-values. Moreover, to enable an unbiased AUC comparison of CNAmet with all other methods (that prioritize all genes), random ranks were assigned to genes labeled by CNAmet with no p-value (non-significant genes). With intCNGEan [20], the weighted Mann-Whitney test with univariate analysis was used with an effective p-value threshold of 0.1. In pint/simcca [34], segmented CN data was used only when the resolution of the CN platform was higher than the resolution of the GE microarray. In PREDA/SODEGIR we used ‘spline’ for smoothing, 1000 random gene orderings of the output regions, and the median AUC as an unbiased output for gene prioritization.

For all methods, each CN probe/segment has been matched to the closest/corresponding GE probe within the same chromosomal arm [34, 38], although the preprocessing of CN data depends partially on the platform resolution [8]. On the latest high-density SNP arrays, for instance, segmentation strategies are essential for estimating the CN for individual genes [8]. Various approaches consider to investigate only certain genomic regions at a time, e.g. to avoid bias, and propose different strategies to select the size of the chromosomal region, including fixed windows in terms of consecutive probes or base pairs [28, 30, 34], chromosome arms or minimal common regions [26] or performing kernel regression [19] where the probe signals are modeled with a smoothing function which accounts for the non-uniform distribution of the genes along the genome.

**Simulated data** Two simulated data sets were generated following Schäfer et al. ([29]; 'Schäfer' data set) and Bicciato et al. ([19]; 'Ferrari' data set). For the 'Schäfer' data set, CN and GE values are drawn from a normal mixture where two components represent aberrations of different extent for each locus; 100 samples were created for each input with mixture proportions of either 10% or 90% for the affected and normal regions. Varying noise levels were imposed using multiple variance parameters (0.25, 0.5, 1, 2 and 4 times an adjusted median absolute deviation of the data). The data points are organized in 16 equally sized blocks to mimic affected regions. The 'Ferrari' data with 6 samples was created by manipulating a renal cell carcinoma data set through permutation of loci and adding or subtracting constants to both CN and GE values within 10 blocks of 10 Mbp. Normal control data was generated by subtracting the median across the samples [19].

**Real case studies** Benchmarking on real case studies is crucial, but defining the ground truth is more challenging than in simulation studies. We investigated two breast cancer data sets [12-13] and a leukemia study [39] using expert-curated lists of known breast cancer genes [40], and leukemia genes from the Cancer Gene Census [41] as the ground truth for the comparisons, respectively. The ‘Hyman’ data set [13] contains 14 breast cancer cell lines, preprocessed as in [3]. The 'Pollack' data set [12] contains 41 breast cancer samples. The 'Mullighan' data set consists of 171 acute lymphoblastic leukemia (ALL) samples divided in 9 subtypes [39, 42]. CN data (Affymetrix Human Mapping 500K) was downloaded from ftp://ftp.stjude.org and normalized with CRMA v2 [43]. The log-additive model from the CRMA v1 algorithm [44] was used for probe summarization. Data values from Nsp and Sty array of the 500K array set were combined and segmented with CBS [45]. Gene expression profiles of the same ALL specimens, measured with the Affymetrix HG-U133A platform were obtained from GEO (GSE12995; [46] and preprocessed with the RPA algorithm (R/Bioconductor; [47]) and EntrezID-based custom chip definition file (v13; [48]). The reference for GE and CN data was defined as the median normalized log-ratios across all samples. Probes with no EntrezID or location information and probes mapping to multiple locations or in sex chromosomes were excluded. GE and CN probes were matched by selecting for each gene the closest CN probe. Missing values were imputed by Gaussian random samples using the mean and variance of the data.

| Implementation | CN preprocessing | Methodological character | Significance scoring | Reference |
|---|---|---|---|---|
| CNAmet (R) | called | custom statistic; two-step | PPT; aberrant regions | Hautaniemi et al., 2004; Louhimo and Hautaniemi 2011 |
| DR-Correlate / t-test (BC) | raw/segmented | two-step | PPT; p-values | Salari et al. 2010 |
| DR-Correlate (BC) / Pearson, Spearman | raw/segmented | COR | PPT; p-values | Salari et al. 2010 |
| edira (R) | raw/segmented | custom statistic; COR | NT; p-values | Schäfer et al. 2009 |
| intCNGEan (R) | cghCall object | custom statistic; two-step | PNT; p-values | van Wieringen et al. 2009 |
| Ortiz-Rivas (R) | raw/segmented | two-step | PNT; p-values | Ortiz et al. 2011 |
| PMA (CRAN) | raw/segmented | LV; COR | PLV; p-values | Witten et al. 2009 |
| PREDA (BC) | raw/segmented | custom statistic; two-step | PPT; aberrant regions/ q-values | Bicciato et al. 2009; Ferrari et al., 2011 |
| pint/simcca (BC) | raw/segmented | LV; COR | PLV; custom statistic | Lahti et al. 2009 |
| SIM (BC) | raw/segmented | REG | PT; p-values | Menezes et al. 2009 |

**Table 1** Summary of the comparison algorithms. The implementations are available through Bioconductor (BC); CRAN; or R source code (R). The copy number preprocessing methods required by each algorithm are listed. Abbreviations: Correlation analysis (COR), regression analysis (REG), latent variables analysis (LV), parametric test (PT), nonparametric test (NT), permutation test based on statistic of nonparametric test (PNT), permutation test based on statistic of parametric test (PPT), permutation test based on latent variable score (PLV).

## Results

The overall cancer gene prioritization performance for the complete gene lists quantified by the AUC analysis is summarized in Figure 1 (for the ROC curves, see Supplementary Figure S1). The highest median ranking across the five benchmarking data sets was obtained by edira (1), followed by Ortiz-Estevez (4) and pint/simcca (4). of these three methods outperformed the other methods on at least one data set. Note that the performance of edira regarding the 'Schäfer' data set and of PREDA/SODEGIR regarding the 'Ferrari' data set needs to be interpreted carefully, because these simulations were originally constructed to follow the particular modeling assumptions of these algorithms in the original publications [19, 29].

Considering the true positive rate among the top-200 genes of each algorithm, pint/simcca had the highest median ranking (1), followed by edira, Ortiz-Estevez and PREDA/SODEGIR (3; Figure 2). These four methods had systematically the highest median rankings with multiple thresholds (20, 50, and 100 top genes); Notably, although edira and PREDA/SODEGIR had the highest scores in the AUC analysis on the Schäfer data, most of other algorithms outperformed these methods with respect to known true positives among the top findings in this data set.

Differences regarding the running times were considerable (Supplementary Table 1). Evaluations were performed on a 64-bit Linux machine with 2 AMD Opteron 2382 processors (K10 architecture, 4 kernels per processor, 2.6 GHz, 32 GB RAM); edira and PMA were the fastest methods with less than one minute running time in all data sets, closely followed by Ortiz-Estevez with a maximum running time of less than 3 minutes. The number of permutations in significance testing affects remarkably the running times of CNAmet, DR-Correlate, intCNGEan and PREDA, although in the latest version of PREDA a parallelized version of the algorithm has been implemented to reduce the computation time [49].

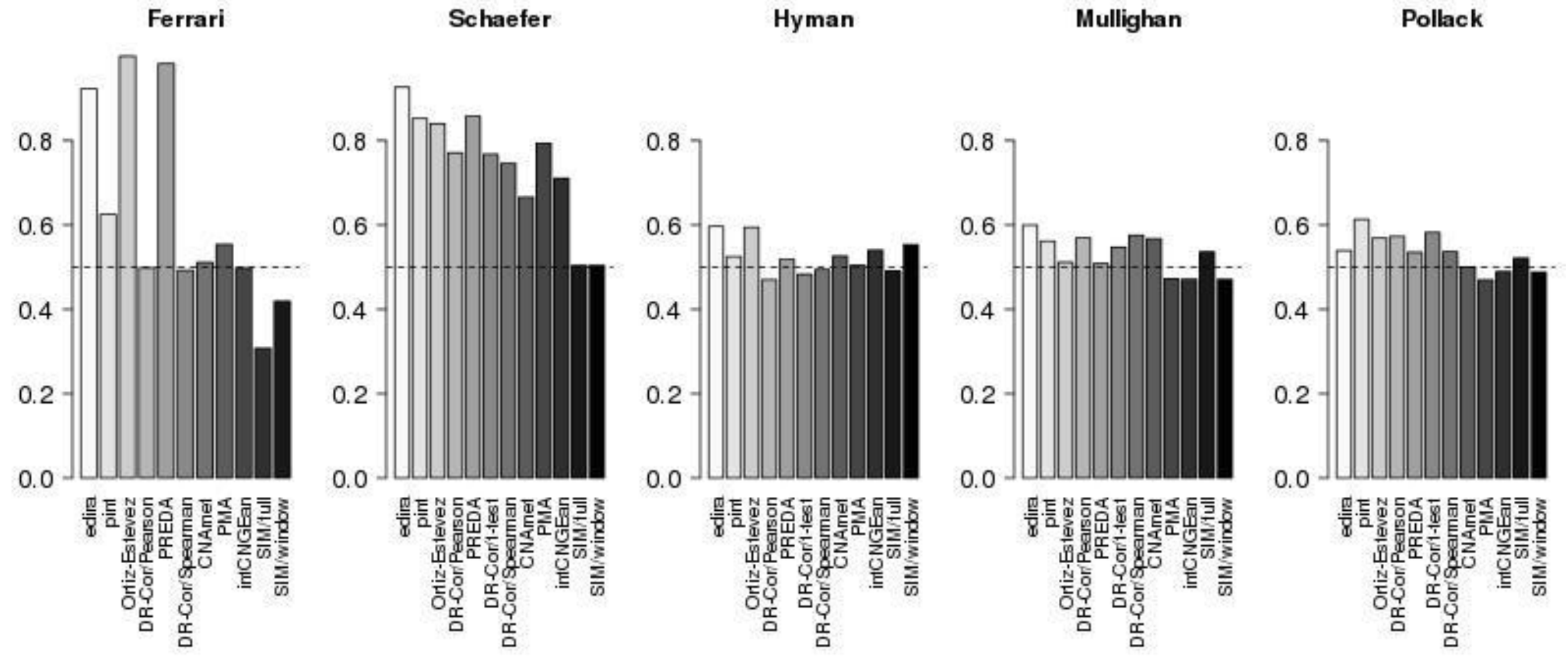

**Figure 1** Area under curve (AUC) values in ROC analysis quantify cancer gene prioritization performance of the methods for the 5 benchmarking data sets. High values indicate high true positive vs. false positive ratio among the top findings; the dashed line indicates the expected AUC value for a random gene list (AUC = 0.5). The methods have been ordered by their median rank across all data sets. For the ROC curves, see Supplementary Figure S1.

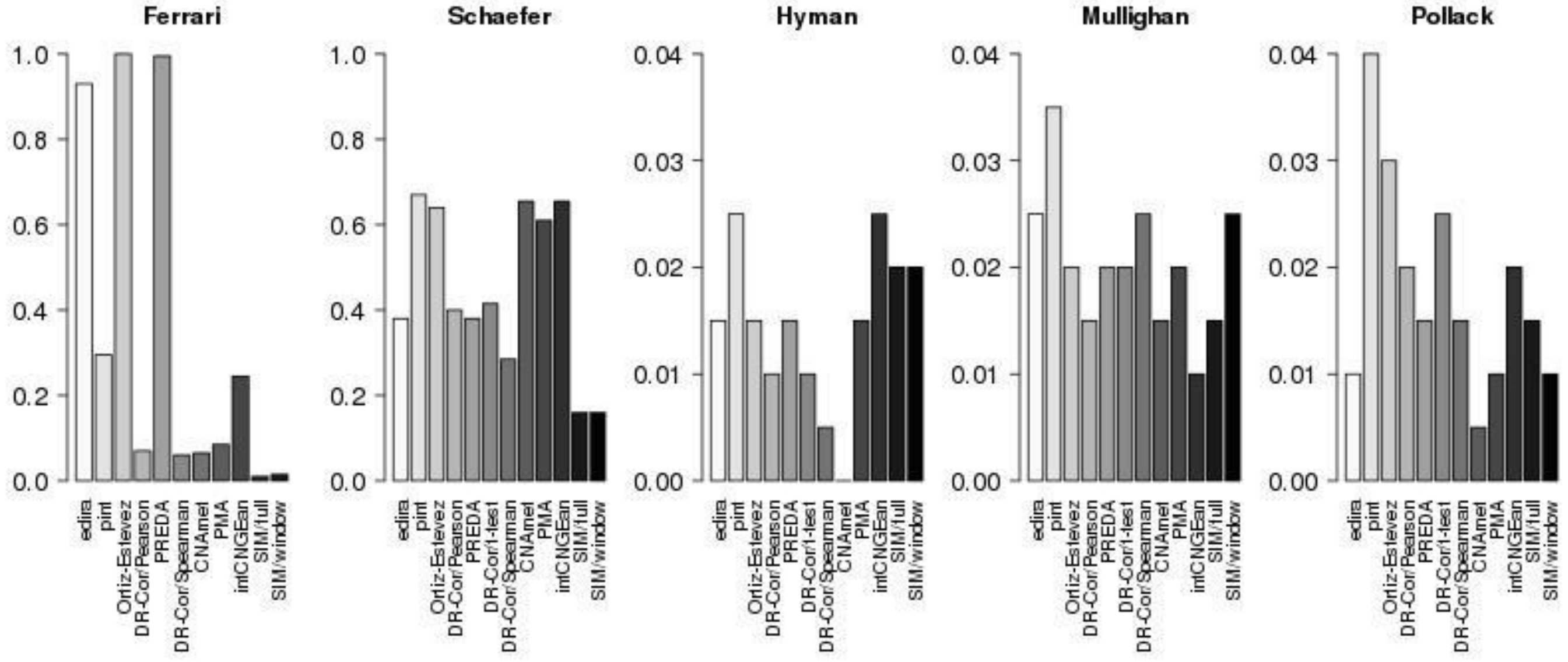

**Figure 2** True positive rates among the top-200 genes from each comparison algorithm across the 5 benchmarking data sets. The overall true positive rate is low in real case studies and the scale for the 'Hyman', 'Mullighan' and 'Pollack' data sets has been accordingly adjusted to highlight the differences. The methods have been ordered as in Figure 1.

## Discussion

Prioritization of disease genes is a key-modeling task in functional genomics [50-53]. This review provides an overview and quantitative benchmarking of publicly available algorithms for detecting associations between GE and CN alterations. This complements the recent review by Huang et al., [8], who pointed out the lack of quantitative comparisons of the available methods. The 'intcomp' benchmarking package applied in this review is freely available at R-forge (http://intcomp.r-forge.r-project.org/), facilitating transparent comparisons and the addition of new algorithms, benchmarking procedures, and validation data sets.

The comparison of 12 different algorithms with respect to their performance in cancer gene prioritization revealed systematic differences between the methods across different data sets, preprocessing scenarios, and sample sizes. The model performance is in general better in the simulation studies, compared to the real cancer data sets, suggesting that manually curated cancer gene lists may be only rough approximations of the ground truth in the real case studies and that the simulations may have lower noise levels. Simulation procedures are only rough approximations of the biological reality and the simulation approach can remarkably affect model performance. For instance, variants of DR-Correlate and CNAmet performed well with 'Schäfer' simulated data but their performance dropped close to random expectation in the 'Ferrari' data set. The 'Ferrari' simulations assume that the CN effect is visible in all tumor samples, which can be particularly disadvantageous for methods that assume high variation from heterogeneous aberration profiles across the samples, including DR-Correlate. The 'Ferrari' and 'Schäfer' simulated data sets were originally introduced in the PREDA/SODEGIR and edira publications which potentially causes positive bias on these methods in the respective data sets. Overall, edira, Ortiz-Estevez, and pint/simcca consistently outperformed the other methods. Considering both relative performance and running time, edira and Ortiz-Estevez seem to offer a good trade-off but all methods have acceptable running times for practical applications.

The choice of preprocessing and model parameters can have a remarkable effect on the results. The key decisions in the context of GE/CN data are associated with selecting the CN preprocessing approach [54], size of the investigated regions, and the matching approach for the integrated data sets. These and related issues are extensively discussed in the recent review by Huang et al. [8]. It is also possible to utilize class information of the samples, for instance by including both tumor and reference samples as in the DR-SAM algorithm [21]. However, in many cases the references are included as a pooled control for two-color microarray experiments but not as a separate group, as with the Hyman and Pollack data sets. Moreover, genomic aberrations often affect only a subset of the cancer patients, and multiple cancer subtypes may be present, as for instance in the Mullighan data set. Further integrative tasks include modeling of trans-regulatory effects of CN aberrations on genes outside the affected region [55-56], disease subtype discovery [4], prediction of patient survival or of clinical covariates [57] and integrative analysis of other data sources such as methylation [58], micro-RNA [59-60], or protein expression [61]. However, fewer implementations for such tasks are currently available. Availability of reference implementations for new computational approaches would facilitate benchmarking and optimizing the algorithms. The standardized benchmarking pipeline introduced in this review can be adjusted to incorporate additional algorithms and data sets as they become available.

## Conclusion

A variety of methods is available for the integrative analysis of gene expression and copy number data. The algorithms can be classified as two-step, regression, correlation-based, and latent variable approaches. Implementation quality, running time and accuracy of the algorithm, as well as preprocessing, sample size and availability of control samples need to be considered when selecting the appropriate computational method. The benchmarking pipeline reveals systematic differences in cancer gene prioritization performance of available implementations across five case studies.

## Key Points

- The integrative analysis algorithms for gene expression and copy number data include two-step, regression, correlation-based, and latent variable approaches
- The benchmarking pipeline reveals systematic differences in the cancer gene prioritization performance of currently available implementations
- Implementation quality, running time and accuracy of the algorithm, as well as data preprocessing, sample size and availability of control samples need to be considered when selecting the analysis approach

## Funding

This work was supported by EuGESMA COST Action BM0801 (European Genomics and Epigenomics Study on MDS and AML). LL has been supported by Helsinki Institute for Information Technology HIIT and Finnish Center of Excellence on Adaptive Informatics Research (AIRC). MD is supported by the European Leukemia Network of Excellence (LSHC-CT-2004); Deutsche Kinderkrebsstiftung [grant number DKS 2010.21] and Carreras Foundation [grant number DJCLS 09/04]. MS is supported by the Deutsche Forschungsgemeinschaft (Research Training Group Statistical Modeling). SB is supported from AIRC Special Program Molecular Clinical Oncology “5 per mille”.

## Acknowledgements

We would like to thank Francesco Ferrari for providing the simulated data set for this study.

# Supplementary Material

| | Ferrari | Schaefer | Hyman | Mullighan | Pollack |
|---|---|---|---|---|---|
| CNAmet | 106.96 | 52.34 | 77.44 | 28.38 | 44.99 |
| DR-Cor/Pearson | 168.50 | 61.11 | 70.26 | 24.39 | 41.41 |
| DR-Cor/Spearman | 310.52 | 120.40 | 135.04 | 45.08 | 78.41 |
| DR-Cor/t-test | - | 67.52 | 68.02 | 26.18 | 43.38 |
| edira | 0.41 | 0.23 | 0.22 | 0.11 | 0.15 |
| intCNGEan | 5.74 | 47.64 | 20.05 | 45.64 | 23.95 |
| Ortiz-Estevez | 0.53 | 2.84 | 0.65 | 1.24 | 1.07 |
| pint | 86.20 | 130.20 | 29.75 | 6.80 | 19.13 |
| PMA | 0.34 | 0.33 | 0.18 | 0.17 | 0.13 |
| PREDA | 79.23 | 155.65 | 59.95 | 360.60 | 106.53 |
| SIM/full | 87.51 | 155.96 | 13.63 | 4.77 | 5.14 |
| SIM/window | 19.15 | 171.96 | 2.81 | 1.28 | 1.40 |

**Supplementary Table 1** Running times (in minutes) for the comparison algorithms in the five benchmarking data sets**.**

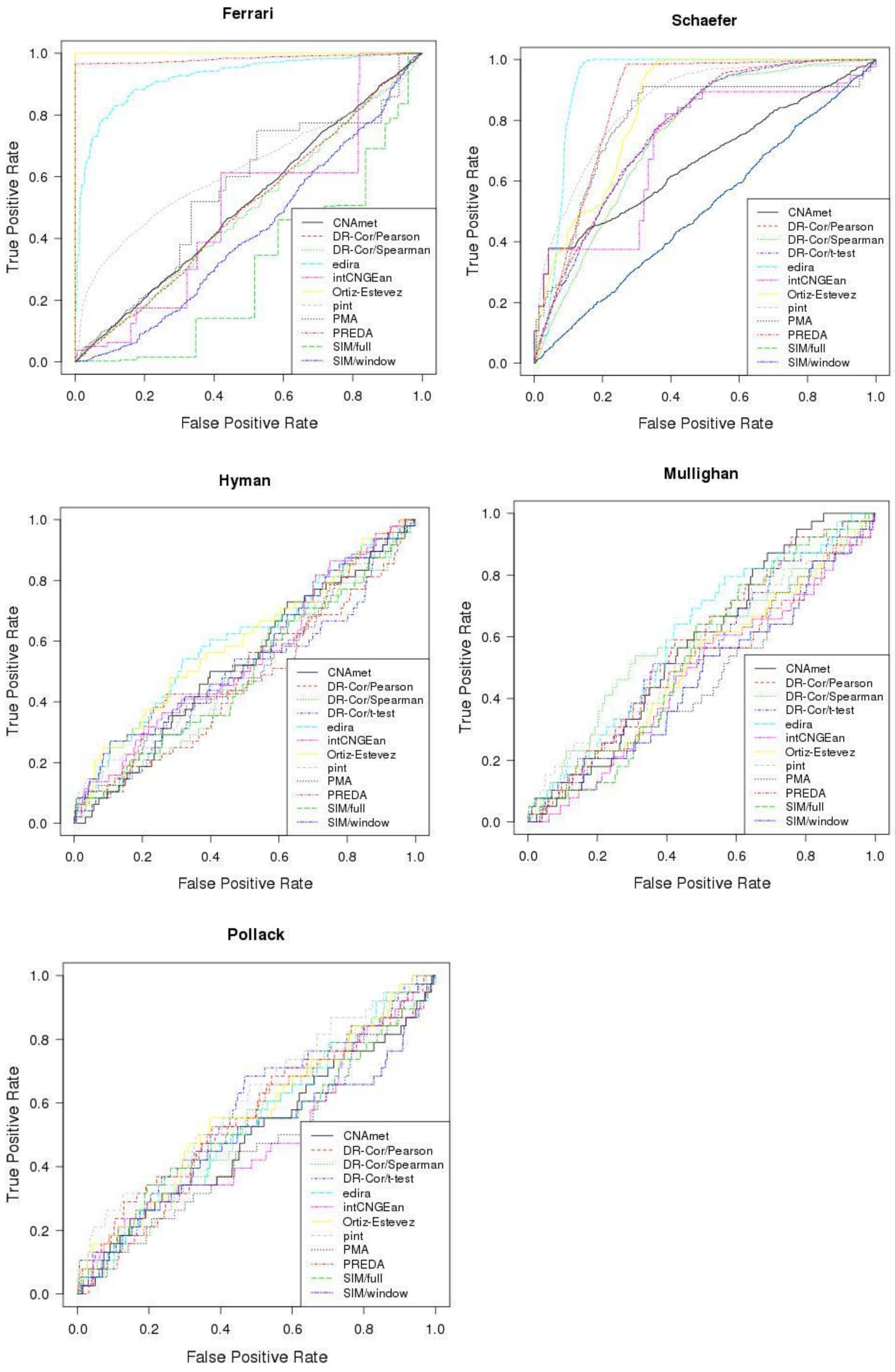

**Supplementary Figure 1** Receiver-Operator Characteristic (ROC) curves characterize the cancer gene prioritization performance of the comparison algorithms in two simulated data sets (‘Ferrari’ and ‘Schäfer’), two breast cancer data sets (‘Hyman’ and ‘Pollack’), and one leukemia data set (‘Mullighan’) based on golden standard lists of custom cancer genes.